# Reconfigurable Parallel Data Flow Architecture

Hamid Reza Naji
International Center for Science and High Technology & Environmental Sciences

*Abstract-* **This paper presents a reconfigurable parallel data flow architecture. This architecture uses the concepts of multi-agent paradigm in reconfigurable hardware systems. The utilization of this new paradigm has the potential to greatly increase the flexibility, efficiency, expandability of data flow systems and to provide an attractive alternative to the current set of disjoint approaches that are currently applied to this problem domain. The ability of methodology to implement data flow type processing with different models is presented in this paper.**

## I. Introduction

The focus of this paper is to illustrate how multi-agent concept can be employed within today's reconfigurable hardware design environments for data flow processing. We call these new agents that run inside reconfigurable logic "Hardware Agents [10]," as opposed to the more traditional software agents that normally reside in computer program memory (RAM) and execute using commercially available microprocessors. Such design environments often utilize hardware description languages such as VHDL to capture the design and use synthesis tools to translate this high level description of the system into a low level bit stream that can be used to program the reconfigurable devices.

We will utilize and adapt a reduced form of the *Belief*, *Desire* and *Intention* (BDI) architecture [9] for our agents. In this architecture, the term, *beliefs*, represent the set of working assumptions that the agent has about itself and the environment in which it functions. This forms informational state of a BDI agent -- where such information may be incomplete and inaccurate but often can be modified in a local manner by the agent as a byproduct of the agent's interactions with other agents and the environment. The term, *desires*, represent the high-level set of objectives and goals that the agent is trying to achieve. The agent's desires must be realistic and must not conflict with each other. *Intentions* represent the deliberative state of the BDI agent. It is here that the detailed sequences of actions, called plans, made to the environment and other cooperating agents through actuators are maintained.

Section 2 introduces the basic concepts associated with the hardware multi-agent paradigm and reconfigurable computing environment. In section 3, paper illustrates the implementing hardware agents for data/control flow type environments in four models: two models that describe deterministic hardware agents in fine and coarse grain modes; one model of hardware agents handling both control flow and data flow; and one intelligent, non-deterministic model that hints at some of the more advanced possibilities of hardware agent use. Section 4 presents implementation and results. The results of implementing dataflow operations with hardware agents show the high processing speed of input tokens and producing results in compare to software agents implementation. Section 5 provides conclusions.

## II. Reconfigurable Hardware Agents

The current state of Field Programmable Gate Array (FPGA) technology and other reconfigurable hardware [13] makes it possible for hardware implementations to enjoy much of the flexibility that was formally only associated with software. Unlike conventional fixed hardware systems, reconfigurable hardware has the potential to be configured in a manner that matches the structure of the application. In some cases, this can be done statically before execution begins where the structure of the hardware will remain unchanged as the system operates. In other cases it is possible to re-configure the hardware dynamically as the systems is operating to allow it to adapt to changes in the environment or the state of the system itself. In other words, the design of the hardware may actually change in response to the demands placed upon the system throughout the scope of the application. The system could be a hybrid of both low-level hardware based agents and higher-level software implementations of agents which cooperate to achieve the desired results. Implementation of agent techniques in re-configurable hardware[11,12] allows for creation of high-speed systems that can exploit a much finer grained parallelism than is possible with distributed software based systems.

It is assumed that an embedded system will be created that utilizes adaptable (reconfigurable) hardware which can be created using FPGA, System on a Chip (SOC)[14], or custom technology. In such an architecture, the functionality of the reconfigurable hardware is controlled by placing design information directly into the configuration memory. In this way, the external environment has the capability to either change the hardware's functionality dynamically or at the time that the application is created by introducing agents into the appropriate area of configuration memory that controls the functionality and interconnectivity of the reconfigurable hardware. In this architecture, the reconfigurable logic is assumed to support partial reconfiguration in that it is assumed that segments of its logic can be changed without affecting other segments (for example the Xilinx Virtex-II architecture supports powerful new configuration modes, including partial reconfiguration. Partial reconfiguration can be performed with and without shutting down the device)[13]. Interaction with the external environment is supported by I/O connections made directly to the reconfigurable logic. This allows high speed sensor and actuator operations to be controlled directly by the reconfigurable logic without processor intervention.

Figure 1 illustrates a generic dynamically adaptable embedded system environments that support the hardware agent model that is proposed in this paper. In Figure 1 an embedded processor/controller is connected to the reconfigurable hardware in a manner that allows it to alter the configuration memory associated with one or more segments of the partially reconfigurable logic. In this model it is the responsibility of the embedded processor to initiate the configuration of each segment of the reconfigurable hardware by transferring configuration data from processor-controlled memory spaces to the configuration memory using memory mapped or DMA type operations. It should also be noted that the configuration memories are shown as if they were spatially separated from the logic elements that they control but this is usually not the case. In general configuration memory is dispersed throughout the reconfigurable hardware.

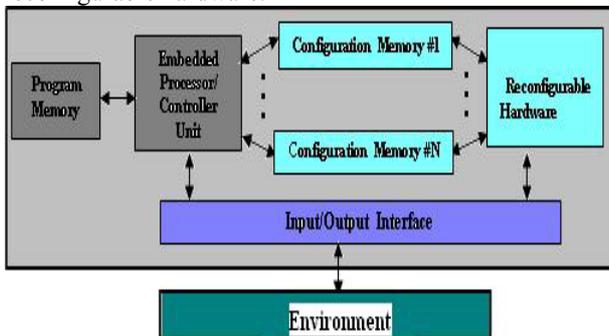

Figure 1. A Processor-Controlled Dynamically Reconfigurable Embedded System Environment

The processing time of a hardware agent can be one or two orders of magnitude greater than an equivalent software agent due to the speed of hardware compared to the speed of microprocessor-based software. This speed could be especially desirable in a real-time system and real-time processing requiring high-speed signal conditioning. In special cases, if the beliefs and the inputs to the agent are expressed as Boolean values, then the function that maps the current set of beliefs and the input data to a new set of beliefs can be implemented as combinatorial logic. The speed of this implementation would be much faster than performing the comparable operation in software. Likewise, if desires and intentions are both expressed as Boolean values, the function that maps desires into intentions can also be implemented in combinatorial logic; again, at very high speed.

### III. Design of Multi Hardware Agent Systems to Implement Data Flow Operations

The ability of hardware agents to implement data flow type synchronization with different models is presented in this section. This type of synchronization is often employed when creating modern hardware to communicate between asynchronous hardware elements. In a data flow operation, the execution of each operation is driven by the data that is available to that operation. The behavior of data flow operations can be shown by data flow graphs(DFG) which represent the data dependencies between a number of operations. A data flow graph is made up of operators (actors) connected by arcs that convey data. An operator has input and output arcs that carry tokens bearing values to and from the actor. When tokens are present on each input arc and there are no tokens on any output arc, actors are enabled (fired). This means removing one token from each input arc, applying the specified operation to the values associated with those tokens, and placing tokens labeled with the result value on the output arcs. We will present four models to show the ability of hardware agents to implement data flow operations in different scenarios.

#### A. Deterministic Fine Grain Hardware Agents

Consider using the dataflow graph shown in figure 2 to find the output $O_1$. In this dataflow graph there are four inputs($I_1, I_2, I_3, I_4$) and 5 nodes(operations).

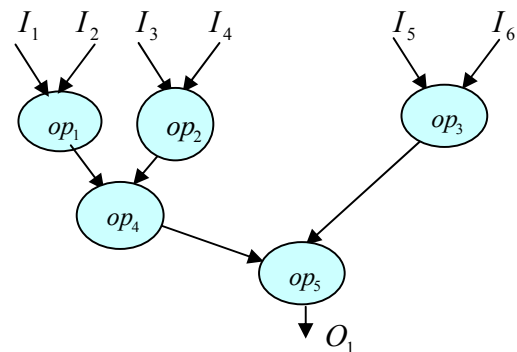

Figure 2. A sample dataflow graph

If we use a multi-agent system to perform this operation, we can implement each of the nodes (operations) with a single agent if we define them at a fine grain level. In this example we use five different agents and each of them runs one single operation as is shown in Figure 3. The agents act in parallel on isolated operations, get information (data) from the environment, and send the results back to the environment.

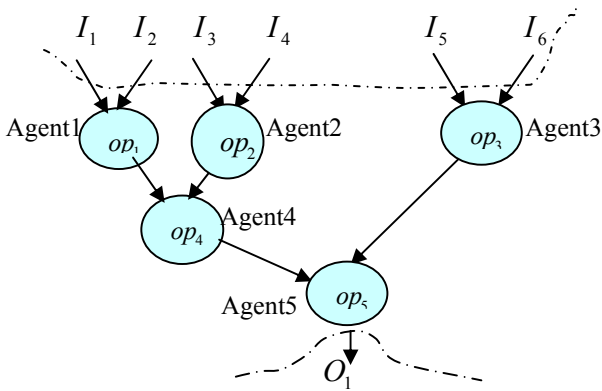

Figure 3. A multi-agent system to implement the data flow operations of Figure 2

By implementing this data flow graph with hardware agents we benefit from the speed of the specialized hardware in processing the input tokens and producing the results. Five hardware agents cooperate together to form a multi-agent architecture for this data flow graph. As Figure 4 shows, in this structure, Agent1, Agent2, and Agent3 receive the input tokens from the enviroment, process them and send the results to Agent4 and Agent5. Finally Agent5 sends the overall result to the environment. A signal from the environment activates this multi-agent system and when each agent completes the operation on its input tokens it will set its done signal and send the value of that result to the agent in the next level. It will inform that agent by sending its done signal to the strobe signal of its successive agent. In this model, hardware agents use done and strobe signals for handshaking.

According to Figure 4, Agent1,Agent2,and Agent3 sends intermediate results through their TR_Agent port and Agent4 and Agent5 receive this information through their RS_Agent port. Agent1, Agent2, Agent3 can use their Request and Acknowledge signals to interact with the environment (send a request which means the agent is free and ready to receive new tokens, and send an acknowledge which means the agent has received the input tokens).

Agent5, after processing the final operation, sends the result through its Output port to the environment and informs the environment by setting its done signal. As we mentioned before, the most important advantage of hardware agents for implementing data flow operations is the the high speed for processing data inputs and producing the outputs. Thus, the speed of information flow can be several times the speed of the flow of information when the same flow graph is implemented in software.

Using the reconfigurability of hardware agents we can reconfigure the agents in the same multi-agent system to implement a different data flow graph. For example, different data flow graphs can be implemented using the same multi-agent system of Figure 4 as we will see later.

In this model agents are small, simple and easy to implement for simple operations, but to implement a complex system we need many agents and a lot of communication between agents with high latency. So, deterministic fine grain hardware agents system is suitable for simple deterministic systems.

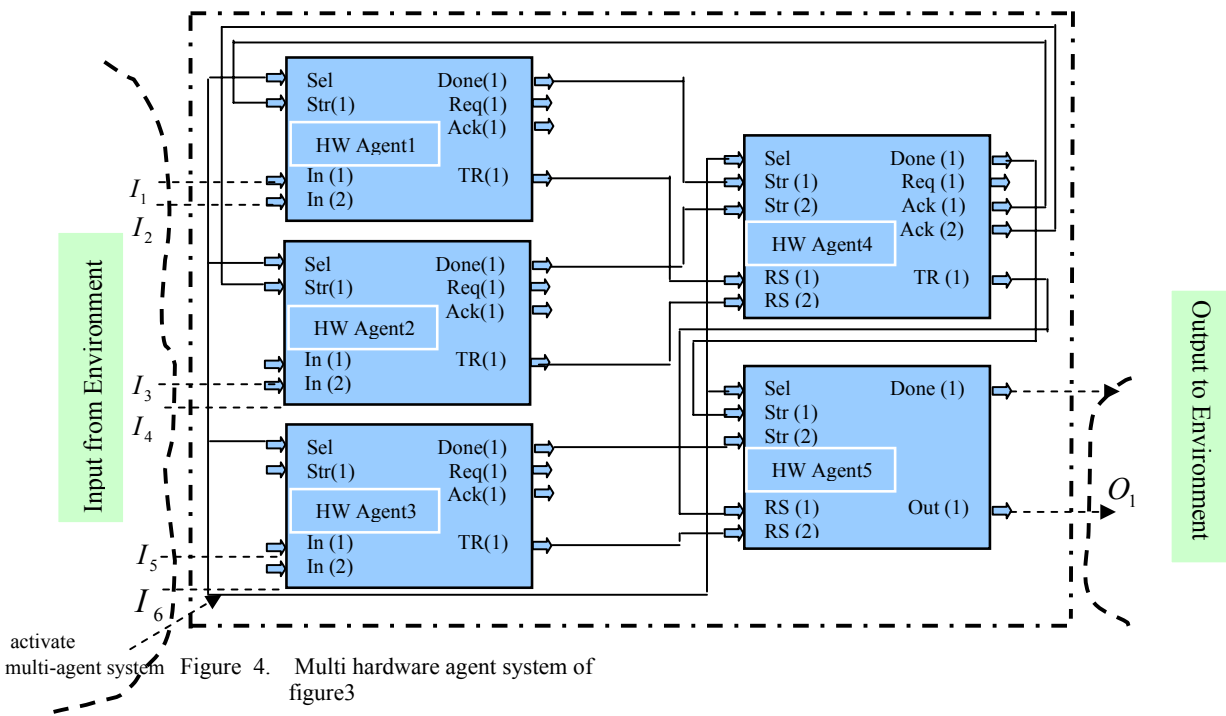

Figure 4. Multi hardware agent system of figure3

## B. Deterministic Fine/Coarse Grain Hardware Agents

To show the power gained by reconfiguring hardware agents and also to demonstrate how hardware agents can provide support for both fine grain and coarse grain abstractions[15], we implement the data flow graph of Figure 5 using the same multi-agent system of Figure 3.

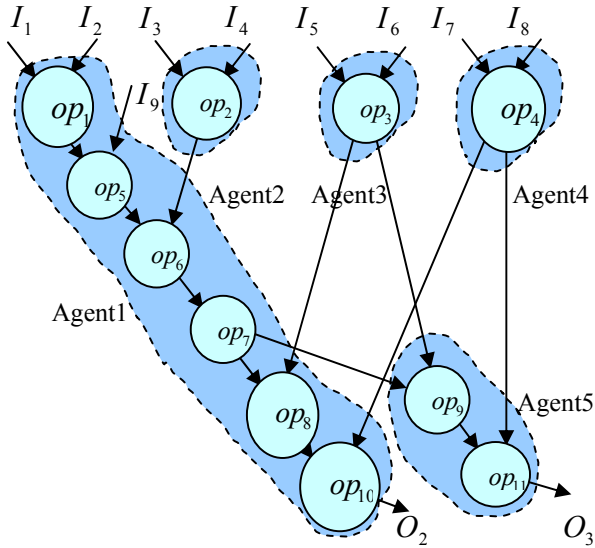

Figure 5. A sample fine grain/coarse grain dataflow graph

As we see in Figure 5, Agent1 and Agent5 are coarse grained agents, with several data flow operations implemented in the same agent,while Agent3, Agent4, and Agent5 are fine grained agents, with only a single data flow operation implemented per agent. They act in parallel in those operations which are not dependent on the other operations, and also cooperate together to find the outputs. The major reason for combining some operations in a single agent is to reduce the hardware interface and the amount of inter-agent communication. This is analogous to the grain packing problem associated with traditional parallel processing problems[16].

We can implement the data flow graph of Figure 5 using the same multi hardware agent system of Figure 4, just agents are reconfigured.

In this model with combination of fine and coarse grain agents a trade of between the agent simplicity and communication time depending to the complexity of system can be provided. So, a combination of fine grain and coarse grain hardware agents is suitable for systems consisting both simple and complex deterministic operations.

## C. Control /Data Flow Hardware Agents

This model demonstrates how control flow as well as data flow can be implemented using hardware agents.We consider the following graph (Figure 6) which contains both control flow and data flow. In this system, according to the events in the environment, the control part will choose the time that the data flow operations $op_1$ & $op_3$ or $op_2$ & $op_4$ should be activated.

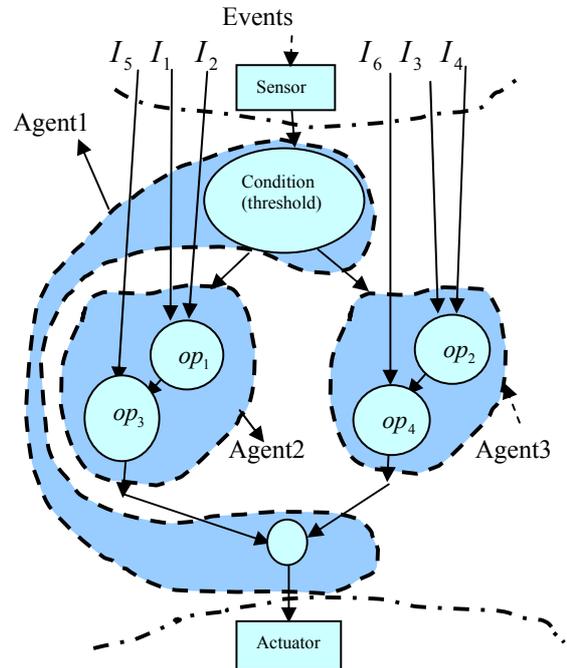

Fig. 6. A sample dataflow graph

We use a multi-agent system with three agents to implement this control and dataflow graph, as shown in Figure 6. In this system, Agent1 is used to implement control flow while Agent2 and Agent3 are used to implement data flow part.

The implementation of this control/data flow graph is with a multi hardware agent system similar to the system shown in Figure 4 but with three hardware agents.

In this model agents can run both data processing and control on the processing in the same multi-agent system at the same time. So, control/data flow hardware agents provides more independent powerful multi-agent systems.

## D. Non-Deterministic Hardware Agents

This model demonstrates how hardware agents can be used in a non-deterministic and intelligent manner. We consider the control and data flow graph shown in Figure 7 that consists of three hardware agents. Agent1 is non-deterministic and intelligent. It receives input information, and saves its current state in memory. Its new state is a combination of the old state, what it has learned from the environment, and what it has calculated itself. In this system, according to the events in the environment, Agent1 will choose to activate Agent2 or Agent3, separately or in tandem, or will choose not to activate them at all. If Agent1 doesn't receive any information within a certain period of time it will timeout and take appropriate action relative to the environment, according to its current state. We can define the learning and decision making capability of Agent1 by the following function description:

Function HW-Agent (*percept*) returns *action*
  Static: *memory* ; the agent's memory
  *memory* ← update-mem(*memory, percept*)
    ;Learning by perception from environment
  *action* ← take-decision(*memory*)
    ;Decision-making by its knowledge
  *memory* ← update-mem(*memory, action*)
    ;Learning by last action
  return *action*

According to this function a hardware agent can learn and update its memory using its current knowledge and percepts (set of perceptions or inputs) from the environment, its current state and calculations based both on state and environmental input. The agent makes decisions using its total knowledge, environmental, state, and current calculations on both environmental and state.

As we see in Figure 7, the possible actions (*plans*) of this multi-agent system, which illustrate its non-determinism are:

Plan1: $op_1 \rightarrow op_3 \rightarrow action$
  *by cooperation of Agent1 & Agent2*

Plan2: $op_2 \rightarrow op_3 \rightarrow action$
  *by cooperation of Agent1 & Agent3*

Plan3: $memory_{agent1} \rightarrow op_3 \rightarrow action$
  *by Agent1(its knowledge & running $op_3$ )*

Plan4: $memory_{agent1} \rightarrow action$
  *by Agent1(its knowlwdge)*

With such a non-deterministic structure, the fault tolerant capability of hardware agents can be easily demonstrated. Suppose Agent1 has a timer, which times out after an input token is not received for a period of time. In this case, a value based on previous state or previous outputs can be presented as the output of the system.

The implementation of this data flow graph is with a multi hardware agent system similar to the system implemented for Figure 6 but with three reconfigured hardware agents.

With considering some situations which are not predefined or cannot predicted mainly in real-time systems then having agents with non-deterministic behavior in the multi-agent system will be useful. So non-deterministic hardware agents provide multi-agent systems with the high capability of responding to the non-deterministic real-time situations.

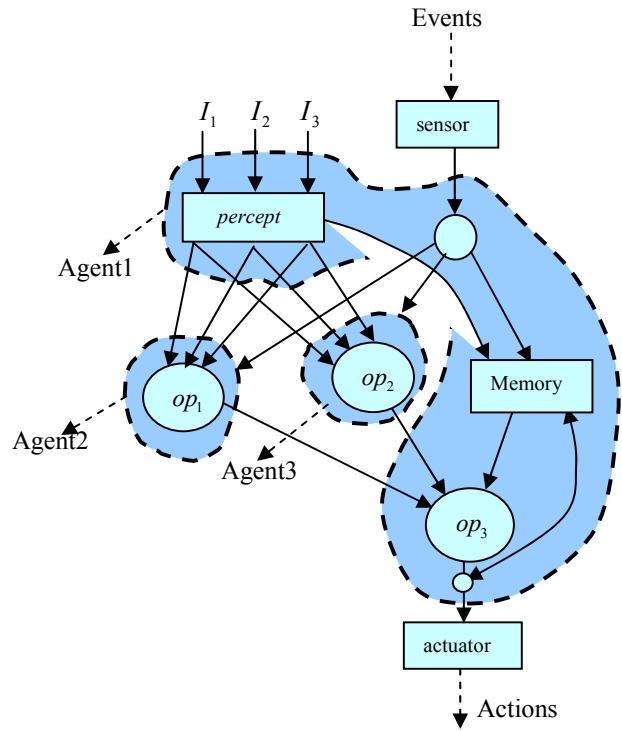

Figure 7. A sample non-deterministic dataflow graph

## IV. Implementation and results

Suppose that we are using dataflow operations of Figure 2, multi-agent system of Figure 3 and multi hardware agents of Figure 4 for a data fusion system as shown in Figure 8. $s_1$ and $s_2$ are the sensory inputs to the system.

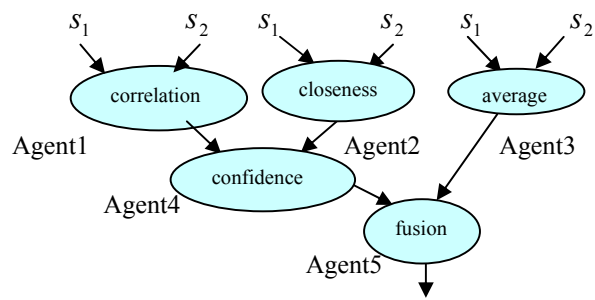

Figure 8. Data flow operations of Figure 3 for data fusion

In the first level of fusion process, Hardware Agent1 computes the correlation between sensors $s_1$ and $s_2$ by using pairs of observation $(a_i, b_i)$ of these sensors using equation 1.

$$\text{Correlation} = \frac{N\sum_1^k a_i b_i - \sum_1^k a_i \sum_1^k b_i}{[N\sum_1^k a_i^2 - (\sum_1^k a_i)^2]^{1/2}[N\sum_1^k b_i^2 - (\sum_1^k b_i)^2]^{1/2}} \quad (1)$$

Thus *Correlation*=1 means perfect correlation, *Correlation*= -1 indicates inverse correlation, and *Correlation*=0 indicates no correlation between the data. In this application it is assumed that inverse correlation is unlikely so that we can use the square of the Correlation, (Correlation)$^2$ to obtain a metric that can be used by advanced stages of the fusion process. Another metric that is computed by Hardware Agent2 is the closeness between two sensors which is defined by equation 2.
Closeness Coefficient:

$$\gamma = 1 - \frac{abs(Sensor_a - Sensor_b)}{Clossness\_threshold} \qquad (2)$$

Here, $Sensor_a$ and $Sensor_b$ are the sensors' values and *Closeness_threshold* is the maximum difference between two sensors. If two sensors have the same value then $\gamma = 1$, otherwise $\gamma$ is less than one.

Hardware Agent3 is calculating the average of sensors' value (a simple fusion method). Hardware Agent4 determines the confidence factor between two sensors which is defined by equation 3.

$Confidence.Factor(a,b) = Min(Correlation(a,b)).\gamma \qquad (3)$

Finally Hardware Agent5 determines the fusion value using the outputs of Agent3 and Agent4.

The code for this model of multi-hardware agent data flow fusion system is written using VHDL. In this model the hardware agents set their initial belief set with the first value of the sensors and thresholds for the correlation and closeness of the sensors' data. They will update their belief set with a new value of its sensor (interaction with the environment) and interaction with the other agents (the main agent will change the correlation and closeness threshold if there is not a high enough degree of confidence of the fusion result). There are several intentions (plans) for this multi-hardware agent system to reach its desire and we assigned each plan to a separate agent to use the collaboration of agents for achieving the global goal or desire which is fusion. The first plan is to determine the correlation between the sensors' data, the second plan is to find the closeness of the sensors' data, the third plan is computing the average of sensors' data, and the forth plan is to find the confidence of system. The desire of this multi hardware agent system is to find the fusion value.

The similar multi software agent system implemented for this model has beliefs, intentions (plans), and desires exactly the same as the hardware agents. It should be noted that if the software agents were implemented in Aglets [17] or a similar software agent framework, that the use of Java and other overhead would make the software agent slower than the version of the software agent that was implemented in C++. This means that our software agents implemented in C++ are more efficient than most traditional software agent implementations.

This would imply that the speed comparison between hardware and software agents that follows is a more strict speed comparison for hardware agents.

The run time of the BDI software agents implemented for this fusion system on a 2.6 GHZ Pentium is 2 us. The run time and speedup of the hardware agent implemented for this fusion system as compared to the equivalent software agent for 8 bit,16 bit, and 32 bit agents for several types of FPGA are shown in Table 1 – Table 3. In 8 and 16 bit modes agents are implemented on Xilinx Virtex II-2v40fg256 and Xilinx- VirtexII-2v10000ff1517 and in 32 bit mode agents are implemented on Xilinx Virtex II-2v500fg456 and Xilinx Virtex II-2v10000ff517.

In each table, the first type of FPGA is the minimum size of the FPGA for each agent and the second type is a common large size FPGA. As the results of these tables show, hardware agents are much faster than the similar software agents for the same application. For example, according to the results of Table 1 and Table 3 the speed of an eight bit hardware agent is 80 times and a thirty two bit hardware agent is 19 times that of a similar software fusion agent, using a Xilinx-Virtex 2v10000ff1517. Of course if the software is, for instance, coded and optimized directly in assembler (a software abstraction level similar than the hardware abstraction level managed by FPGAs synthesis tools), all the software layers present in a general purpose computer such as operating systems procedures removed, and the FPGA re/configuration time is taken into account, then the speed up should be a little bit lower.

Table 1. 8 bits Hardware Agents

| FPGA \ Agent | 8 bits | |
|---|---|---|
| | run time | speedup |
| Xilinx- VirtexII 2v40fg256 | 26ns | 77 |
| Xilinx- VirtexII 2v10000ff1517 | 25 ns | 80 |

Table 2. 16 bits Hardware Agents

| FPGA \ Agent | 16 bits | |
|---|---|---|
| | run time | speedup |
| Xilinx- VirtexII 2v80fg256 | 64 ns | 31 |
| Xilinx- VirtexII 2v10000ff1517 | 51 ns | 39 |

Table 3. 32 bits Hardware Agents

| FPGA \ Agent | 32 bits | |
|---|---|---|
| | run time | speedup |
| Xilinx- VirtexII 2v500fg456 | 113 ns | 17 |
| Xilinx- VirtexII 2v10000ff1517 | 106 ns | 19 |

.

Table 4 – Table 6 show the number of logic gates used in each device for the implementation and utilization of hardware agents. Device selection varies according to the policy of the design (distributed or concentrated) and the size and the number of agents that we need to build our hardware agent system. For example according to the results of Table 4 and Table 6 we can implement up to 18 eight bit hardware agents and up to 4 thirty two bit hardware agents similar to the data flow operation system of Figure 6 in each Xilinx-Virtex 2v10000ff1517.

Table 4. Device Utilization (8 bits HW Agents)

| Resource \ FPGA | Xilinx- VirtexII 2v10000ff1517 | |
|---|---|---|
| | Used | Utilization |
| Ios | 225 | 21.00 % |
| Function Generators | 5314 | 4.43 % |
| CLB Slices | 2658 | 4.41 % |
| DFFs or Latches | 192 | 2.00 % |

Table 5. Device Utilization (16 bits HW Agents)

| Resource \ FPGA | Xilinx- VirtexII 2v10000ff1517 | |
|---|---|---|
| | Used | Utilization |
| Ios | 113 | 10.54 % |
| Function Generators | 2048 | 1.70 % |
| CLB Slices | 1024 | 1.70 % |
| DFFs or Latches | 160 | 1.66 % |

Table 6. Device Utilization (32 bits HW Agents)

| Resource \ FPGA | Xilinx- VirtexII 2v10000ff1517 | |
|---|---|---|
| | Used | Utilization |
| Ios | 57 | 5.32 % |
| Function Generators | 939 | 0.78 % |
| CLB Slices | 471 | 0.77 % |
| DFFs or Latches | 148 | 1.54 % |

### V. Conclusion

In this paper, a general architectural framework for implementing agents in reconfigurable hardware has been presented. Hardware implementations have always been known to be faster than software implementations, but at the cost of great loss in flexibility. The use of reconfigurable hardware added flexibility to hardware, while still retaining most of the speed of hardware. Now the use of hardware agents can greatly expanded this flexibility of reconfigurable hardware. In the future, such improvements to reconfigurable hardware such as faster programming times, and more independently reconfigurable sections in the reconfigurable hardware will make hardware agents even more flexible while coming even closer to retaining the speed that makes hardware-based implementations desirable.

The hardware agents developed for data flow application display many of the features associated with more traditional agents implemented in software. The results of hardware agents implementation in this paper show that the speed of hardware agents can be over an order of magnitude greater than an equivalent software agent implementation. The parallel nature of the reconfigurable hardware would cause this speedup to be further increased if more than one agent were implemented in the reconfigurable hardware. It is believed that the use of hardware agents may prove useful in a number of application domains, where speed, flexibility, and evolutionary design goals are important issues.

### References

[1] Walter B, Zarnekow R. Intelligent Software Agents, Springer-Verlag, Berlin Heidelberg, New York, NY, 1998.
[2] Jennings N, Wooldrige M. Agent Technology, Springer-Verlag, New York, NY, 1998.
[3] Jennings N, Wooldridge M. Intelligent Agents: Theory and Practice, The Knowledge Engineering Review, 1995; 10(2):115-152.
[4 Brooks R. Intelligence Without Reason, Massachusetts Institute of Technology, Artificial Intelligence Laboratory, A.I. Memo, 1991.
[5] Ambrosio J, Darr T. Hierarchical Concurrent Engineering in a Multi-agent Framework, Concurrent Engineering Research and Application Journal, 1996;4:47-57.
[6] Weiss G. Multiagent Systems-A Modern Approach to Distributed Artificial intelligence, Cambridge: MIT Press 1999.
[7] Flores-Mendez R. Towards a Standardization of Multi-agent System Frameworks, ACM Crossroads— Special Issue on Intelligence Agents, 1999;5(4):18-24.
[8] Jennings N, Sycara K, Wooldridge M. A Roadmap of Agent Research and Development, Autonomous Agents and Multi-Agent Systems Journal, Kluwer Publishers, 1998;1(1):7-38.
[9] Rao A. BDI Agents: From Theory to Practice, ICMAS '95 First International Conference on Multi-agent System, 1995.
[10] Naji H. R., Wells B. E., Aborizka M., Hardware Agents, Proceedings of the ISCA 11[th] International Conference on Intelligent Systems on Emerging Technologies (ICIS-2002), Boston, MA, 2002.
[11] Naji H. R., Implementing data flow operations with multi hardware agent systems, Proceedings of the IEEE 2003 Southeastern Symposium on System Theory, Morgantown, WV, March 2003.
[12] Naji H. R., Wells B.E., Etzkorn L., Creating an Adaptive Embedded System by Applying Multi-agent

.

Techniques to Reconfigurable Hardware, Future Generation Computer Systems, 2004, (20) 1055-1081.

[13] Guccione S. Reconfigurable Computing at Xilinx, Proceedings of Euromicro Symposium on Digital Systems Design, 2001.

[14] Becker J, Pionteck T, Glesner M. Adaptive Systems-on-chip: Architectures, Technologies and Applications, 14th Symposium on Integrated Circuits and Systems Design, 2001.

[15] Srinivasan V, Govindarajan S, Vemuri R. Fine-Grained and Coarse-grained behavioral partitioning with effective utilization of memory and design space exploratin for multi-FPGA architecture, IEEE Transactions on very large scale integration (VLSI) systems, 2001.

[16] Kruatrachue B, Lewis T. Grain Size Determination for Parallel Processing, IEEE trans. On Software, 1998;5(1):23-32.

[17] G. Karjoth, D.B. Lange, A security Model for Aglets, Internet Comput, IEEE,1997;1(4):68-77.

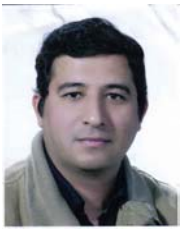

Hamid Reza Naji is an assistant professor in the International Center for Science and High Technology & Environmental Sciences in Kerman, Iran. His research interests include embedded, reconfigurable, and multiagent systems, networks, and security. Naji has a PhD in computer engineering from the University of Alabama in Huntsville, USA. He is a professional member of the IEEE. Contact him at hamidnaji@ieee.org

.